\documentclass[twocolumn,10pt]{elsart}
\usepackage{epsfig}
\usepackage{xspace}
\journal{Astroparticle Physics}

\def\modell{{\sl poly-gonato} model\xspace}

\def\knee{{\sl knee}\xspace}
\def\knees{{\sl knees}\xspace}

\def\gcm2{g/cm$^2$}

\def\eref#1{eq.\,(\ref{#1})}
\def\fref#1{Fig.\,\ref{#1}}

\def\rref#1{Ref.\,\cite{#1}}
\def\sref#1{Sect.\,\ref{#1}}

\begin{document} 
\runauthor{J.R. H\"orandel, N.N. Kalmykov, A.V. Timokhin}
\begin{frontmatter}
\title{Propagation of super high-energy cosmic rays in the Galaxy}
\author[uka]{J\"org R. H\"orandel},
\author[si]{Nikolai N. Kalmykov}, and
\author[lm]{Aleksei V. Timokhin}
\address[uka]{Institute for Experimental Nuclear Physics, University of
         Karlsruhe, P.O. Box 3640, 76021 Karlsruhe, Germany}
\address[si]{Skobeltsyn Institute of Nuclear Physics, Lomonosov Moscow State
         University, Leninskie Gory 1, Moscow, 119992 Russia}
\address[lm]{Faculty of Physics, Lomonosov Moscow State University, Russia}

\begin{abstract}
 The propagation of high-energy cosmic rays in the Galaxy is investigated.
 Solutions of a diffusion model are combined with numerically calculated
 trajectories of particles. The resulting escape path length and interaction
 path length are presented and energy spectra obtained at Earth are discussed.
 It is shown that the energy spectra for heavy elements should be flatter as
 compared to light ones due to nuclear interactions during the propagation
 process. The obtained propagation properties of ultra-heavy elements indicate
 that these elements could play an important role for the explanation of the
 second knee in the cosmic-ray energy spectrum around 400~PeV.
\end{abstract}

\end{frontmatter}

\section{Introduction}
The explanation of the origin of super-high energy cosmic rays is one of the
unsolved problems in astrophysics. The energy spectra at the sources are not
identical to the observed spectra at Earth. Hence, studying the sources is
closely related to investigations of cosmic-ray propagation processes in the
Galaxy.  For the latter, a detailed knowledge of the structure of the magnetic
fields is important. Unfortunately, the configuration of the galactic magnetic
field remains an open question --- different models can explain the
experimental data \cite{berezinsky,ruzmaikin,ptuskin,gorchakov}.  

How cosmic rays are accelerated to extremely high energies is another unsolved
problem.  Although the popular model of cosmic-ray acceleration by shock waves
in the expanding shells of supernovae (see e.g.
\cite{ellison,berezhko,sveshnikova}) is almost accepted as "standard theory,"
there are still serious difficulties.  Furthermore, the question about other
acceleration mechanisms is not quite solved, and such mechanisms could lead to
different cosmic-ray energy spectra at the sources \cite{berezinsky}.  

Different concepts can be verified, calculating the primary cosmic-ray energy
spectrum, making assumptions on the density of cosmic-ray sources, the energy
spectrum at the sources, and the configuration of the galactic magnetic fields.
The diffusion model may be used in the energy range $E<10^{17}$~eV, where the
energy spectrum is obtained using the diffusion equation for the density of
cosmic rays in the Galaxy. At higher energies this model ceases to be valid,
and it becomes necessary to carry out numerical calculations of particle
trajectories for the propagation in the magnetic fields.  This method works
best for the highest energy particles, since the time required for the
calculations is inversely proportional to the particle energy.

Therefore, a calculation of the cosmic-ray spectrum in the energy range
$10^{14}-10^{19}$~eV has been performed in a combined approach: solutions of a
diffusion model are used at low energies and particle trajectories are
numerically integrated at high energies.

In \sref{assect} the basic assumptions for the diffusion model will be
described. The results obtained with the propagation model are presented in the
subsequent sections.  The calculated propagation path length and interaction
probability of cosmic rays will be discussed in \sref{pathsect} and
\sref{intersect}, respectively. Finally, the energy spectra are presented in
\sref{specsect}, followed by a discussion of the results (\sref{discussion}).

\section{Assumptions} \label{assect}
High isotropy and a comparatively long retention of cosmic rays in the Galaxy
($\sim10^7$ years for the disk model) reveal the diffusion nature of particle
motion in the interstellar magnetic fields. This process is described by
a corresponding diffusion tensor \cite{berezinsky,ptuskin,kalmykov}.  The
steady-state diffusion equation for the cosmic-ray density $N(r)$ is
(neglecting nuclear interactions and energy losses)
\begin{equation} \label{diffeq}
 - \nabla_iD_{ij}(r)\nabla_jN(r)=Q(r) .
\end{equation}
$Q(r)$ is the cosmic-ray source term and $D_{ij}(r)$ the diffusion tensor.

Under the assumption of azimuthal symmetry and taking into account the
predominance of the toroidal component of the magnetic field, \eref{diffeq} is
presented in cylindrical coordinates as 
\begin{eqnarray} \label{zylfun}
\left[ 
  -\frac{1}{r}\frac{\partial}{\partial r} r D_\perp \frac{\partial}{\partial r} 
  -           \frac{\partial}{\partial z}   D_\perp \frac{\partial}{\partial z} 
  -           \frac{\partial}{\partial z}   D_A     \frac{\partial}{\partial r} 
  \right.  \nonumber \\ \left.
  +\frac{1}{r}\frac{\partial}{\partial r} r D_A     \frac{\partial}{\partial z} 
\right] N(r,z)=Q(r,z) ,
\end{eqnarray}
where $N(r,z)$ is the cosmic-ray density averaged over the large-scale
fluctuations with a characteristic scale $L\sim100$~pc \cite{ptuskin}.
$D_\perp\propto E^m$ is the diffusion coefficient, where $m$ is much less than
one ($m\approx0.2$), and $D_A\propto E$ the Hall diffusion coefficient.  The
influence of Hall diffusion becomes predominant at high energies
($>10^{15}$~eV). The sharp enhancement of the diffusion coefficient leads to an
excessive cosmic-ray leakage from the Galaxy at energies $E>10^{17}$~eV.  To
investigate the cosmic-ray propagation at such energies it becomes necessary to
calculate the trajectories for individual particles.  

Such a numerical calculation of trajectories is based on the solution of the
equation of motion for a charged particle in the magnetic field.
The calculation was carried out using a fourth-order Runge-Kutta method.
Trajectories of cosmic rays were calculated until they left the Galaxy.
Testing the differential scheme used, it was found that the accuracy of the
obtained trajectories for protons with an energy $E=10^{15}$~eV after passing a
distance of 1~pc amounts to $5\cdot10^{-8}$~pc.  The retention time of a proton
with such an energy averages to about 10 million years, hence, the total error
for the trajectory approximation by the differential scheme is about 0.5~pc.

The magnetic field of the Galaxy consists of a large-scale regular and a
chaotic, irregular  component $\vec{B}=\vec{B}_{reg}+\vec{B}_{irr}$.  A purely
azimuthal magnetic field was assumed for the regular field 
\begin{eqnarray}
 B_z=0, \quad B_r=0, \nonumber \\ B_\phi=1~\mu\mbox{G} \cdot 
       \exp\left(-\frac{z^2}{z_0^2}-\frac{r^2}{r_0^2}\right) ,
\end{eqnarray}
where $z_0=5$~kpc and $r_0=10$~kpc are constants \cite{ptuskin}. These values
are adopted from \rref{ptuskin} to ensure the same conditions for both methods,
i.e. trajectory calculations and the diffusion approach.  The irregular field
was constructed according to an algorithm used in \cite{zirakashvili}, that
takes into account the correlation of the magnetic field intensities in
adjacent cells.  The radius of the Galaxy is assumed to be 15~kpc and the
galactic disk has a half-thickness of 200~pc. The position of the Solar system
was defined at $r=8.5$~kpc, $\phi=0^\circ$, and $z=0$~kpc. A radial
distribution of supernovae remnants along the galactic disk was considered as
sources \cite{kodaira}.

\section{Propagation Path Length} \label{pathsect}

\begin{figure}[t]
  \includegraphics[width=\columnwidth]{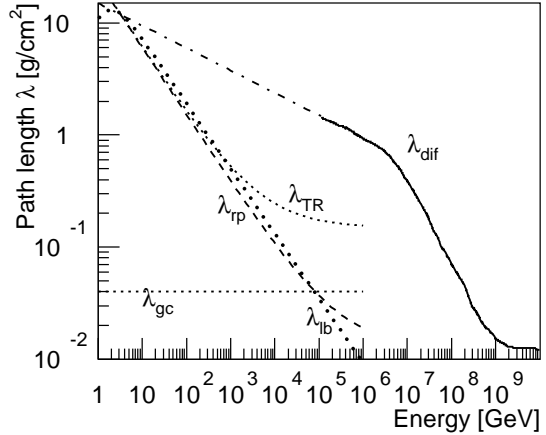}
  \caption{\label{pathlength} Path length in the Galaxy for protons.
	   The values for the diffusion model ($\lambda_{dif}$) are indicated
	   by the solid line. They are extrapolated to lower energies by the
	   dashed dotted line. Also shown are predictions of a leaky-box model
	   ($\lambda_{lb}$, \eref{leakyboxfun}), a residual path length model
	   ($\lambda_{rp}$, \eref{swordyfun}), and an upper limit for a
	   residual path length model according to the TRACER experiment
	   ($\lambda_{TR}$) \cite{tracer05}.  The horizontal line indicates the
	   matter to be passed along a straight line from the galactic center
	   to the solar system ($\lambda_{gc}$).}
\end{figure}

Assuming an interstellar matter density $n_d=1$~cm$^{-3}$ for the galactic disk
and $n_h=0.01$~cm$^{-3}$ for the halo, following Chap. 3 in \rref{berezinsky},
trajectory calculations were performed at energies above 0.1~PeV.  The
dependence of the path length on energy was obtained from the dependence of the
transport time for protons in the galactic disk and the halo.  The resulting
escape path length for protons as function of energy is presented in
\fref{pathlength} as $\lambda_{dif}$.  For heavier nuclei with charge $Z$ the
path length scales with rigidity, i.e. is related to the values for protons
$\lambda^p(E)$ as $\lambda(E,Z)=\lambda^p(E/Z)$. 

For protons at 4~PeV, the amount of traversed material is approximately
0.7~\gcm2.  At higher energies, the calculated path length decreases as
$\propto E^{-0.7}$.  Between 0.1 and 1~PeV the calculations yield a behaviour
$\lambda\propto  E^{-\delta}$ with $\delta=0.2$.  The dashed dotted line
indicates a trend below 0.1~PeV extrapolating the calculated values to lower
energies using the slope obtained. This yields a path length around 15~\gcm2 at
1~GeV, well compatible with measured values, see below.  The relatively small
$\delta=0.2$ is much lower than the value usually assumed in Leaky-Box models
($\delta\approx0.6$). On the other hand, such a slope of $\delta=0.2$ follows
from the diffusion model proposed in \rref{ptuskin}.

Measurements of the ratio of secondary to primary cosmic-ray nuclei at GeV
energies are successfully described using a leaky-box model. For example,
assuming the escape path length for particles with rigidity $R$ and velocity
$\beta=v/c$ as
\begin{equation} \label{leakyboxfun}
 \lambda_{lb}=\frac{26.7\beta ~\mbox{\gcm2}}
              {\left(\frac{\beta R}{1.0~{\rm GV}}\right)^{0.58} + 
	       \left(\frac{\beta R}{1.4~{\rm GV}}\right)^{-1.4}}
\end{equation}
various secondary to primary ratios, like boron to carbon, phosphorus to
sulfur, and sub-iron to iron, as obtained by the ACE/CRIS and HEAO-3
experiments, can be described consistently in the energy range from
$\sim70$~MeV to $\sim30$~GeV \cite{cris-time}.
A similar approach is the residual path length model \cite{swordy}, assuming the
relation
\begin{equation} \label{swordyfun}
 \lambda_{rp} = \left[ 6.0 \cdot 
     \left(\frac{R}{10~{\rm GV}}\right)^{-0.6} + 0.013 \right] 
     \frac{\rm g}{{\rm cm}^2}
\end{equation}
for the escape path length.  Recent measurements of the TRACER experiment yield
an upper limit for the constant term in the residual path length model of
0.15~\gcm2 \cite{tracer05}.  The three examples are compared to the predictions
of the diffusion model in \fref{pathlength}.

Extrapolating these relations to higher energies leads to extremely small
values at PeV energies due to the strong dependence of the path length on
energy ($\propto E^{-0.6}$). For the approaches according to \eref{leakyboxfun}
and \eref{swordyfun} above $10^5$~GeV the traversed matter would be less than
the matter passed along a straight line from the galactic center to the solar
system $\lambda_{gc}= 8~\mbox{kpc}\cdot 1~\mbox{proton/cm}^3 \approx
0.04$~\gcm2. This value is indicated in \fref{pathlength} as dotted line.  For
the upper limit obtained by TRACER the path length at $10^5$~GeV amounts to
about 0.17~\gcm2.  This value does not exclude diffusion, as the larmor radius
$r_L=1.08~\mbox{pc}\cdot E [\mbox{PeV}]/(Z\cdot B[\mu\mbox{G}])$ of a proton
with an energy of 0.1~PeV in the galactic magnetic field ($B=1~\mu$G) is about
0.1~pc and 0.17~\gcm2 correspond to about 34~pc or $340r_L$ at 0.1~PeV. The
value may be even larger when considering cosmic-ray propagation in the
galactic halo.  However, in this approach the cosmic-ray sources are expected
to be closer to the solar system as in the diffusion model, since the latter
yields a path length of about 1.5~\gcm2 at 0.1~PeV. This value follows from the
assumption that the sources are disributed like supernova remnants and, hence,
the question may arise whether this assumption is true.

\begin{figure}[t]
  \includegraphics[width=\columnwidth]{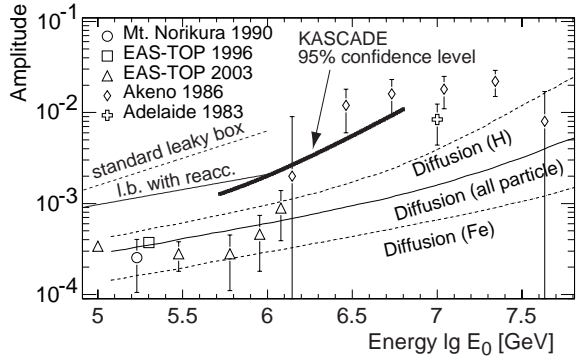}
  \caption{\label{aniso}Rayleigh amplitudes as function of energy for various
	   experiments, taken from \rref{kascade-aniso}.  The results obtained
	   by Mt. Norikura \cite{norikura-aniso}, EAS-TOP
	   \cite{eastop-aniso96,eastop-aniso03}, Akeno \cite{akeno-aniso},
	   Adelaide \cite{adelaide-aniso}, and KASCADE \cite{kascade-aniso} are
	   compared to predictions of leaky-box models \cite{ptuskinaniso} and
	   a diffusion model \cite{candiaaniso}.  For the diffusion model,
	   predictions for primary protons, iron nuclei, and all particles are
	   displayed.}
\end{figure}

At low energies (GeV regime) the ratio of primary to secondary nuclei is
measured to derive information about cosmic-ray propagation. At higher energies
in the PeV domain such information is experimentally not accessible and other
information is used to set constraints on propagation models.  One possibility
is the anisotropy of the arrival direction of cosmic rays.  It can be
characterized by Rayleigh amplitudes.  Values observed by different experiments
are compiled in \fref{aniso} \cite{kascade-aniso}.  They are compared to
predictions of propagation models.  Two versions of a leaky-box model
\cite{ptuskinaniso}, with and without reacceleration are represented in the
figure.  Leaky-box models, with their extremely steep decrease of the path
length as function of energy ($\lambda\propto E^{-0.6}$), yield relative large
anisotropies even at modest energies below $10^6$~GeV, which seem to be ruled
out by the measurements, see also \cite{ptuskinaniso}.  The measured values are
almost an order of magnitude smaller.  On the other hand, a diffusion model
\cite{candiaaniso}, which is based on the same basic idea \cite{ptuskin} as the
present work, predicts relatively small values at low energies and a modest
rise only. In the figure, predictions for pure protons and iron nuclei are
shown together with calculations for a mixed composition. The predicted
Rayleigh amplitudes are compatible with the measured values. This may indicate,
that diffusion models are a more realistic description of cosmic-ray
propagation in the Galaxy at PeV energies.

\section{Interaction Probability} \label{intersect}

\begin{figure}[t]
  \includegraphics[width=\columnwidth]{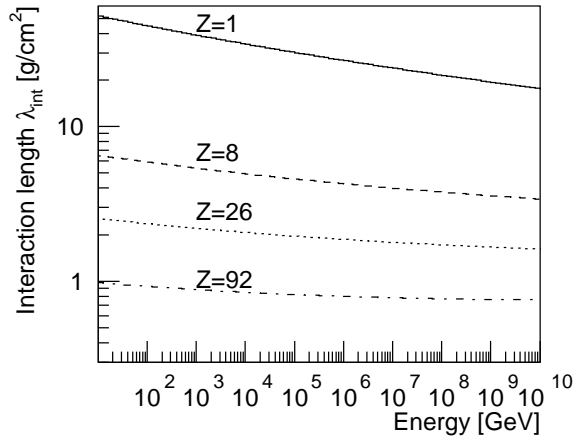}
  \caption{\label{lambdaint}Interaction length as function of energy for
	   different elements based on cross-sections according to the hadronic
	   interaction model QGSJET. The nuclear charge numbers are indicated.}
\end{figure}

To calculate the interaction length of nuclei in the interstellar medium,
nuclear cross sections according to the hadronic interaction model QGSJET
\cite{qgsjet} have been used.
The cross sections for nuclei with mass number $A>1$ have been parameterized
using
\begin{equation}
 \sigma(E)[\mbox{mb}]=\alpha(E)A^{\beta(E)} ,
\end{equation}
where $\alpha$ and $\beta$ are
approximated as
\begin{eqnarray}
 \alpha(E) = 50.44 -7.93 \lg\left(\frac{E}{\mbox{eV}}\right) \nonumber \\
              + 0.61 \lg^2\left(\frac{E}{\rm eV}\right)
\end{eqnarray}
and
\begin{equation}
 \beta(E)= 0.97 -0.022 \lg\left(\frac{E}{\rm eV}\right)	.
\end{equation}
The corresponding interaction lengths for four different elements are presented
in \fref{lambdaint}.  The values decrease slightly as function of energy.
Values for protons are in the range $55-20$~\gcm2, the values decrease as
function of nuclear charge and reach values $<1$~\gcm2 at all energies for the
heaviest elements.

\begin{figure}[t]
  \includegraphics[width=\columnwidth]{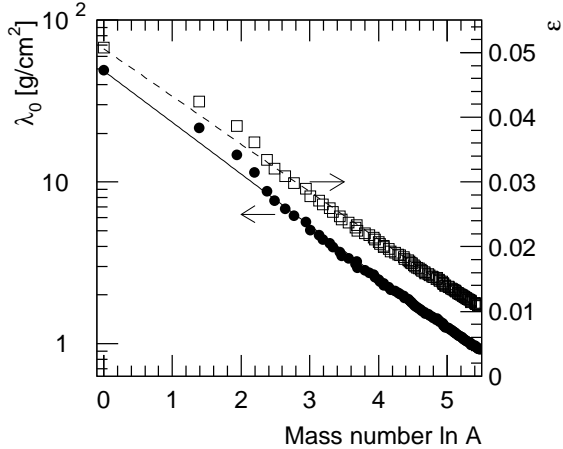}
  \caption{\label{lambdaslope}
	   Parameters $\lambda_0$ at 10~GeV (filled symbols, left scale) and
	   $\varepsilon$ (open symbols, right scale) to describe the
	   interaction length as function of energy according to
	   \eref{lambdainteq} for nuclei with mass number $A$.}
\end{figure}

The dependence of the interaction length on energy has been approximated by the
relation
\begin{equation} \label{lambdainteq}
 \lambda_{int}=\lambda_0 \left(\frac{E}{10~{\rm GeV}}\right)^{-\varepsilon} .
\end{equation}
The corresponding values for all elements are depicted in \fref{lambdaslope} as
function of the logarithm of the mass number $A$.  The values for $\lambda_0$
at 10~GeV decrease from about 50~\gcm2 for hydrogen to about 0.9~\gcm2 for
uranium.  The small energy dependence is reflected in the values for
$\varepsilon$ ranging from 0.05 for protons to 0.01 for the heaviest element.
The lines illustrate the approximately linear dependence of $\lg \lambda_0$ and
$\varepsilon$ on $\ln A$.  Small deviations from this behavior are visible for
the light elements helium, lithium, beryllium, and boron.  The small values for
$\varepsilon$ motivate why $\lambda_{int}$ is frequently assumed to be
constant.

\begin{figure}[t]
  \includegraphics[width=\columnwidth]{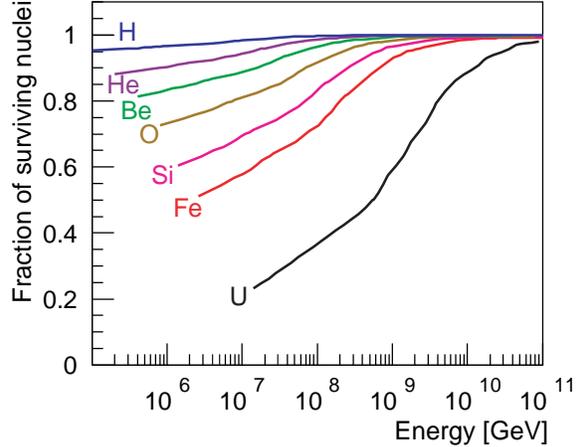}
   \caption{\label{fragment}Fraction of nuclei surviving without interaction in
            the Galaxy for different elements as function of energy.}
\end{figure}

Using the path length $\lambda_{dif}$ and the interaction length obtained, the
interaction probability for different nuclei has been calculated.  Nuclear
fragmentation is taken into account in an approximate approach \cite{fragm}. It
should be pointed out that a nuclear fragment conserves the trajectory
direction of its parent if $Z/A$ in question is the same as for the primary
nucleus and for most stable nuclei the ratio $Z/A$ is close to $1/2$. The
resulting fraction of nuclei which survive without an interaction is presented
in \fref{fragment} for selected elements. 

In the energy range shown protons reach the Earth almost undisturbed without
any interaction, less than 5\% of them interact.  On the other hand, for
heavier elements nuclear interactions play an important role and a significant
fraction of them interacts at low energies.  However, it turns out that even
for ultra-heavy elements (heavier than iron) at the expected \knees (at an
energy of approximately $Z \cdot 4.5$~PeV), see below, more than about 50\% of
the nuclei survive without interactions.  This is an important result, which
could help to understand the origin of the second \knee in the all-particle
spectrum around 400~PeV, as shall be discussed below.  It should be noted that
the fraction of surviving nuclei is even larger for a leaky-box model, with its
low path length at such high energies.

\section{Energy Spectra} \label{specsect}

\begin{figure}[t]
  \includegraphics[width=\columnwidth]{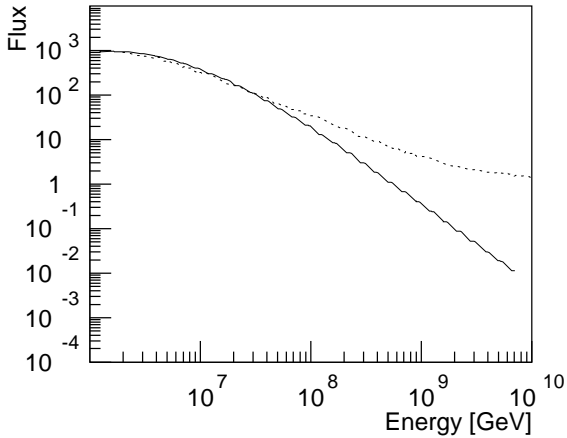}
  \caption{\label{pspec}Calculated spectra of protons for the diffusion model
	   (solid line) and the numerical trajectory calculations (dotted
	   line).}
\end{figure}

The results for the calculations of the cosmic-ray proton spectrum are
presented in \fref{pspec}. They were obtained using the diffusion model and
numerical calculations of trajectories. It is evident from the graph that both
methods give identical results up to about $3\cdot10^7$~GeV. At higher energies
there is a continuous decrease of the intensity in the diffusion spectrum,
which has its origin in the excessive increase in the diffusion coefficient
that leads to large leakage of particles from the Galaxy.  

An energy of $10^8$~GeV can be taken as the conventional boundary to apply the
diffusion model. At this energy the results obtained with the two methods
differ by a factor of two and for higher energies the diffusion approximation
becomes invalid.

Although the \knee in the all-particle spectrum has been observed more than 40
years ago \cite{kulikov}, it was only recently that experimental spectra for
groups of elements became available.  The KASCADE air shower experiment derived
energy spectra for five groups of elements, namely protons, helium, CNO,
silicon group, and iron group \cite{ulrichapp}.  The spectra exhibit a fall-off
for individual elements at high energies. 
These results and the data available from other experiments are compatible with
the \modell \cite{pg}, assuming a \knee for each element at an energy of about
$Z\cdot 4.5$~PeV \cite{aspenreview}.

\begin{figure}[t]
  \includegraphics[width=\columnwidth]{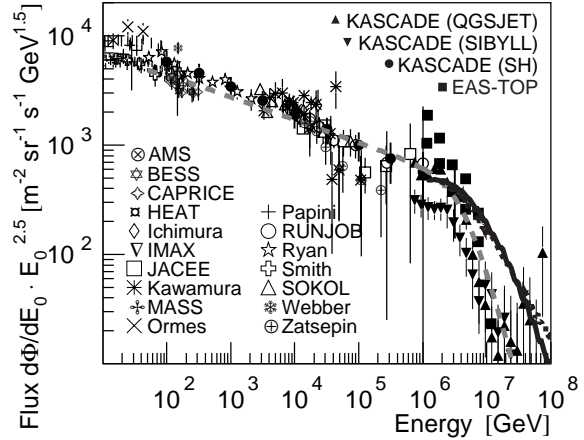}
  \caption{\label{pflux} Proton flux as obtained from various measurements, for
	   references see \cite{vulcano}, compared to the spectra shown
	   in \fref{pspec} (black lines) and the \modell \cite{pg} (grey,
	   dashed line).}
\end{figure}

In the following, we compare the predicted spectra already shown in
\fref{pspec} to direct and indirect measurements of the primary proton flux in
\fref{pflux}.  The predicted spectra are normalized to average experimental
values at 1~PeV.  In the range depicted, almost no difference is seen between
the two approaches. The relatively steep decrease of the measured flux at
energies exceeding 4~PeV is not reflected by the predictions. On the other
hand, the data are described reasonably well by the \modell \cite{pg}, shown in
the figure as well.  The observed change in the spectral index
$\Delta\gamma\approx2.1$ according to the \modell has to be compared to the
value predicted by the diffusion model. In the latter the change should be
$1-m\approx0.8$ \cite{ptuskin}. The observed value is obviously larger, which
implies that the remaining change of the spectral shape should be caused by a
change of the spectrum at the source, e.g.  due to the maximum energy attained
in the acceleration process.

The maximum energy and, therefore, the energy at which the spectrum steepens
depends on the intensity of the magnetic fields in the acceleration zone and on
a number of assumptions for the feedback of cosmic rays to the shock front.
The uncertainty of the parameters yields variations in the maximum energy
predicted by different models up to a factor of 100 \cite{berezhko,origin}.  It
seems, there is no consensus about what the "standard model" is considered to
predict. For the time being, it is difficult to draw definite conclusions from
the comparison between the experimental spectra for different elemental groups
and the "standard model" of cosmic-ray acceleration at ultra high energies.  

\section{Discussion} \label{discussion} 

The energy spectra for individual elements measured at the Earth with GeV and
TeV energies can be described by power laws $dN/dE\propto E^\gamma$ with values
for the spectral index $\gamma$ in the range -2.46 to -2.95 for elements from
hydrogen to nickel \cite{wiebel,pg}.  The measurements seem to indicate that
the steepness of the energy spectra at Earth depends on the mass of the nuclei,
heavier elements seem to have flatter spectra.  At higher energies in the PeV
domain the measured spectra are compatible with the assumption of a \knee for
individual elements at about $Z\cdot 4.5$~PeV \cite{pg,aspenreview}. 

The energy spectrum of cosmic rays at their source $Q(E)$ is related to the
observed values at Earth $N(E)$ as
\begin{equation} \label{sourceeq}
 N(E) \propto Q(E) 
  \left( \frac{1}{\lambda_{esc}(E)} + \frac{1}{\lambda_{int}(E)}\right)^{-1}
\end{equation}
with the escape path length $\lambda_{esc}$ and the interaction length
$\lambda_{int}$. Values for the former are presented in \fref{pathlength} and
for the latter in \fref{lambdaint}. The relation between $N(E)$ and $Q(E)$ is
governed by the absolute values of $\lambda_{esc}$ and $\lambda_{int}$ as well
as their respective energy dependence.  The interaction length $\lambda_{int}$
is almost independent of energy, the values for $\varepsilon$ in
\eref{lambdainteq} are smaller than 0.05, see \fref{lambdaslope}. On the other
hand, the propagation path length $\lambda_{esc}$ decreases as function of
energy as $\lambda_{esc}\propto E^{-\delta}$, with values between $\delta=0.6$
for Leaky Box models and $\delta=0.2$ for the diffusion model described in this
work (see \sref{pathsect}).  

In the "standard picture" of galactic cosmic rays usually
$\lambda_{int}>\lambda_{esc}$ is assumed with an energy independent interaction
length and an escape path length $\lambda_{esc}=\lambda_{lb}\propto E^{-0.6}$.
Thus, the relation between $N(E)$ and $Q(E)$ is dominated by the energy
dependence of the escape path length. To explain the spectrum observed at Earth
$N(E)\propto E^{-2.7}$ the spectrum at the sources has to be $Q(E)\propto
E^{-2.1}$.

Recent measurements by the HESS experiment of the TeV gamma ray flux from the
shell type supernova remnant RX J1713.7-3946 yield a spectral index
$\gamma=-2.19\pm0.09\pm0.15$ \cite{hesssnr}.  For the first time, spectra have
been derived directly at a potential cosmic-ray source.  The results are
compatible with a nonlinear kinetic theory of cosmic-ray acceleration in
supernova remnants and imply that this supernova remnant is an effective source
of nuclear cosmic rays, where about 10\% of the mechanical explosion energy are
converted into nuclear cosmic rays \cite{voelkrxj1713}. 

These investigations show that the spectrum at the source is $Q(E)\propto
E^{-2.2}$, a value relatively close to the "standard model". The escape path
length required to obtain the spectra observed at Earth should be $\propto
E^{-0.5}$, i.e. $\delta$ is 0.1 smaller than in the "standard picture".  Two
questions remain open:  Do all potential cosmic-ray sources exhibit a similar
spectrum? And, what fraction of galactic cosmic rays is accelerated in
supernova remnants with spectra like RX J1713?  

\begin{figure*}[t] \centering
  \includegraphics[width=0.9\textwidth]{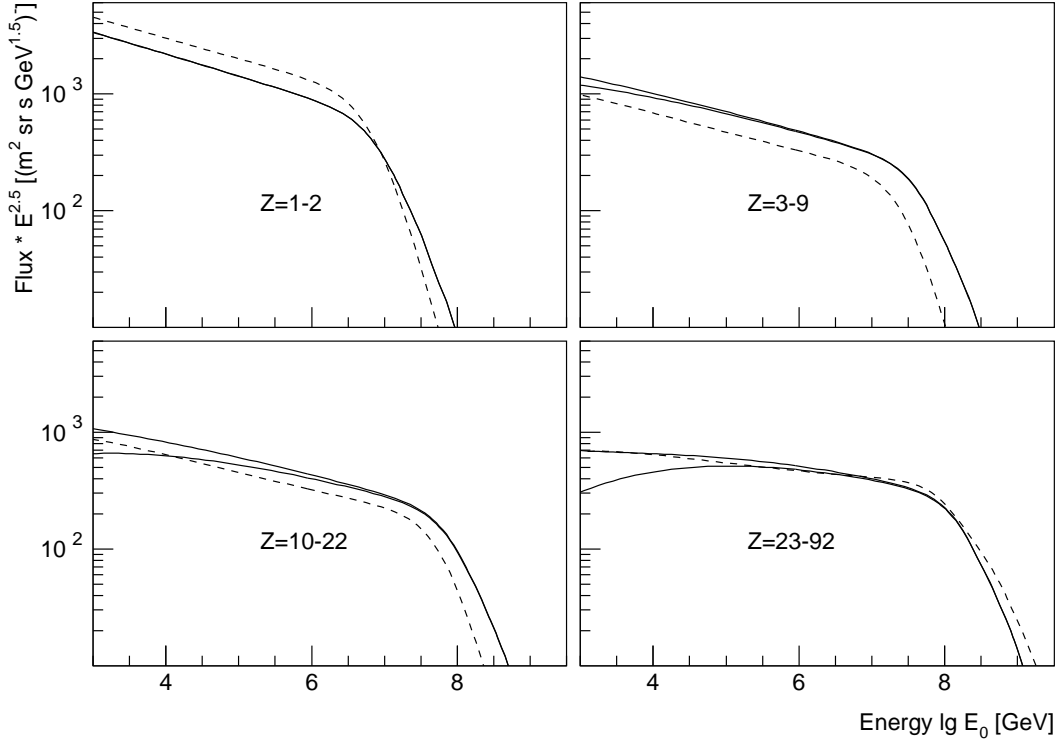}
   \caption{Energy spectra at Earth for elements with nuclear charge $Z$ as
	   indicated. The dashed lines represent spectra according to the
	   \modell, the solid lines are expected from the diffusion model
	   discussed, see text.  Two solid lines are shown in each panel,
	   representing an estimate for the uncertainties.  \label{prop}}
\end{figure*}

Considering the spectra of nuclei observed at Earth, the second term in
\eref{sourceeq} has to be taken into account.  The values of $\lambda_{int}$
for protons are at all energies larger than the escape path length
$\lambda_{dif}$. Hence, for protons the escape from the Galaxy is the dominant
process influencing the shape of the observed energy spectrum.  For iron nuclei
at low energies hadronic interactions are dominating
($\lambda_{int}\approx2.5~\mbox{\gcm2} < \lambda_{dif}$) and leakage from the
Galaxy becomes important only at energies approaching the iron \knee ($26\cdot
\hat{E}_p$).  Finally, for elements heavier than iron, the interaction path
length is smaller than the escape path length at all energies, except above the
respective \knees. For these elements the propagation path length exceeds the
interaction path length by about an order of magnitude at low energies.  Hence,
nuclear interaction processes are dominant for the shape of the observed
spectrum.  Taking this effect into account, it is expected that the energy
spectra for heavy elements should be flatter as compared to light nuclei.  The
spectral index for iron nuclei should be about 0.13 smaller than the
corresponding value for protons.  Direct measurements indicate indeed that the
spectra of light elements are flatter as compared to heavy elements \cite{pg},
e.g. the values for protons $\gamma_p=2.71$ and iron $\gamma_{Fe}=2.59$ differ
as expected. This more or less obvious effect has been pointed out earlier in
several articles, see e.g. \cite{biermann,jones}. However, in the present work
we extend the considerations to a wider energy range and ultra-heavy nuclei
using contemporary estimates of the cross sections.

The \modell extrapolates results from direct measurements in the GeV and TeV
regime to higher energies \cite{pg}. In the model fall-offs for individual
elements proportional to the nuclear charge are assumed at energies of
$E_Z=Z\cdot \hat E_p$, with the value $\hat E_p=4.5$~PeV for protons. The \knee
is caused by the subsequential fall-offs for all elements, starting with
protons. The second \knee in the all-particle spectrum around 400~PeV is
interpreted as the end of the galactic component, where the heaviest elements
(actinides) reach their fall-off energies ($92\cdot \hat E_p\approx414$~PeV).
At this energy, elements heavier than iron are expected to contribute with
about 40\% to the all-particle flux \cite{aspenphen}.  The model uses an
empirical relation for the spectral index. The spectral slope decreases as
function of nuclear charge.  In the following, this behavior is investigated
quantitatively with the propagation model discussed above.

As source composition, the abundances of elements from hydrogen to uranium as
measured in the solar system \cite{cameron} have been weighted with $Z^{3.2}$.
This choice is arbitrary to a certain extent, but may be motivated by a higher
efficiency in the injection or acceleration processes for nuclei with high
charge numbers. The abundances are scaled with a factor which is identical for
all elements to obtain approximately the absolute values as expected at the
Earth according to the \modell.  At the source, a power law $\propto E^{-2.5}$
has been assumed for all elements with a \knee, caused e.g.  by the maximum
energy attained during the acceleration, at $Z\cdot4.5$~PeV, with a power law
index $-3.5$ above the respective \knee.  Using the derived propagation path
length and interaction length, the amount of interacting particles has been
determined.  Secondary products generated in spallation processes are taken
into account, assuming that the energy per nucleon is conserved in these
reactions. They are added to the corresponding spectra with smaller $Z$. The
spectra thus obtained are compared to spectra according to the \modell in
\fref{prop}.  

Two features should be noted: The absolute fluxes at Earth are predicted quite
well, especially when considering that only a simple scaling law has been
introduced for the abundances at the sources, starting with the composition in
the solar system.  More important for the present discussion is the shape of
the spectra. As expected, the shape of the proton spectrum is not influenced by
the (few) interactions during propagation and the difference of the spectral
index at the source and at Earth $\gamma=-2.71$ \cite{pg} can be explained by
the energy dependence of the escape path length $\propto E^{-0.2}$.  On the
other hand, it can be recognized that due to nuclear interactions the spectra
for heavier elements are indeed flatter. The slopes obtained with the simple
approach for the CNO, silicon, and iron groups agree well with the steepness as
expected from the \modell.  For heavy elements at low energies secondary
products generated in spallation processes play an important role for the shape
of the spectrum. At low energies many nuclei interact due to the large escape
path length and the small interaction length, thus, the spectra of nuclei
without any interaction deviate from power laws. However, the spallation
products of heavier elements at higher energies compensate the effect and the
resulting spectra are again approximately power laws, as can be seen in
\fref{prop}.

\section{Summary}
The results obtained show the effectiveness of the combined method to calculate
the cosmic-ray spectrum using a numerical calculation of trajectories and a
diffusion approximation.  The propagation path length in the Galaxy and the
energy spectra at Earth have been calculated to investigate the effect of
escape from the Galaxy as possible origin of the \knee in the cosmic-ray energy
spectrum.  The calculated dependence of the propagation path length on energy
suggests difficulties with the "standard model" of galactic cosmic rays.  An
energy dependence of the propagation path length at low energies $\propto
E^{-0.2}$ is expected, which requires a spectrum at the sources $\propto
E^{-2.5}$ in order to explain the observed spectra at Earth.  To explain the
relatively steep fall-off of the observed energy spectra for elemental groups
at their respective \knees, the modulation of the spectrum due to propagation
solely is not sufficient. An additional steepening of the spectra at the source
is necessary, e.g. caused by the maximum energy attained during acceleration.
It can be concluded that the \knee in the energy spectrum of cosmic rays has
its origin most likely in both, acceleration and propagation processes.

In the \modell the \knee in the energy spectrum at $\hat E_p=4.5$~PeV is caused
by a cut-off of the light elements and the spectrum above the \knee is
determined by the subsequent cut-offs of all heavier elements at energies
proportional to their nuclear charge number. The second \knee around
$400~\mbox{PeV} \approx 92\cdot\hat{E}_p$ is due to the cut-off of the heaviest
elements in galactic cosmic rays. Considering the calculated escape path length
and nuclear interaction length within the diffusion model, it seems to be
reasonable that the spectra for heavy elements are flatter as compared to light
elements. The calculations show also that even for the heaviest elements at the
respective \knee energies more than about 50\% of the nuclei survive the
propagation process without interactions. This may explain why ultra-heavy
elements could contribute significantly ($\sim 40\%$) to the all-particle flux
at energies around 400~PeV and thus explain the second knee in the energy
spectrum.

\section*{Acknowledgments}
The authors are grateful to J.~Engler, A.I.~Pavlov, and V.N.~Zirakashvili for
useful discussions.
N.N.K. and A.V.T. acknowledge the support of the RFBR (grant 05-02-16401).


\end{document}